\documentclass[manuscript]{emulateapj}

\newcommand{\csr}{Chem. Soc. Rev.}
\newcommand{\jms}{J. Mol. Spectr.}
\newcommand{\jphysb}{J. Phys. B: Mol. Opt. Phys.}
\newcommand{\jmst}{J. Mol. Struct.}

\newcommand{\jphys}{J. Physique}
\newcommand{\nature}{Nature}

\shorttitle{Confirmation of circumstellar PH$_3$}
\shortauthors{Ag\'undez et al.}

\begin{document} 

\title{Confirmation of circumstellar phosphine}

\author{M. Ag\'undez\altaffilmark{1}, J. Cernicharo\altaffilmark{1}, L. Decin\altaffilmark{2,3}, P. Encrenaz\altaffilmark{4}, and D. Teyssier\altaffilmark{5}}

\altaffiltext{1}{Instituto de Ciencia de Materiales de Madrid, CSIC, C/ Sor Juana In\'es de la Cruz 3, 28049 Cantoblanco, Spain}

\altaffiltext{2}{Sterrenkundig Instituut Anton Pannekoek, University of Amsterdam, Science Park 904, NL-1098, Amsterdam, The Netherlands}

\altaffiltext{3}{Instituut voor Sterrenkunde, Katholieke Universiteit Leuven, Celestijnenlaan 200D, 3001 Leuven, Belgium}

\altaffiltext{4}{LERMA, Observatoire de Paris, 61 Av. de l'Observatoire, 75014 Paris, France}

\altaffiltext{5}{European Space Astronomy Centre, Urb. Villafranca del Castillo, PO BOX 50727, 28080 Madrid, Spain}

\begin{abstract}

Phosphine (PH$_3$) was tentatively identified a few years ago in the carbon star envelopes IRC~+10216 and CRL~2688 from observations of an emission line at 266.9 GHz attributable to the $J=1-0$ rotational transition. We report the detection of the $J=2-1$ rotational transition of PH$_3$ in IRC~+10216 using the HIFI instrument on board {\emph Herschel}\thanks{Herschel is an ESA space observatory with science instruments provided by European-led Principal Investigator consortia and with important participation from NASA.}, which definitively confirms the identification of PH$_3$. Radiative transfer calculations indicate that infrared pumping to excited vibrational states plays an important role in the excitation of PH$_3$ in the envelope of IRC~+10216, and that the observed lines are consistent with phosphine being formed anywhere between the star and 100 $R_*$ from the star, with an abundance of 10$^{-8}$ relative to H$_2$. The detection of PH$_3$ challenges chemical models, none of which offers a satisfactory formation scenario. Although PH$_3$ locks just 2 \% of the total available phosphorus in IRC~+10216, it is together with HCP, one of the major gas phase carriers of phosphorus in the inner circumstellar layers, suggesting that it could be also an important phosphorus species in other astronomical environments. This is the first unambiguous detection of PH$_3$ outside the solar system, and a further step towards a better understanding of the chemistry of phosphorus in space.

\end{abstract}


\section{Introduction}

Phosphorus being a biogenic element, the study of the various forms in which it is present in the different astronomical environments is a matter of great interest, although still poorly understood \citep{macia2005}. Only five P-containing molecules (PN, CP, HCP, PO, and C$_2$P) have been observed in the gas phase of interstellar and circumstellar media \citep{tur1987,ziu1987,gue1990,agu2007,ten2007,hal2008}. These five molecules are all observed in circumstellar envelopes around evolved stars. However, in interstellar clouds only PN is observed and with a relatively low abundance, implying that most of phosphorus must be elsewhere, probably depleted in dust grains \citep{tur1990}.

Phosphine (PH$_3$), the phosphorus cousin of ammonia, is a relatively stable molecule that could be locking an important fraction of phosphorus in various astronomical environments. Since more than 30 years ago, PH$_3$ is known to be present in the atmospheres of the giant gaseous planets Jupiter and Saturn, where it is the major phosphorus carrier \citep{bre1975,lar1977,wei1996}. Upper limits obtained in Neptune and Uranus imply that the gas phase abundance of phosphorus is probably subsolar in the atmosphere of these icy giants \citep{mor2009}. Phosphine ice is also a plausible major phosphorus constituent of comets, although recent searches in the gaseous coma of a few comets turned out unsuccessful, providing upper limits not significant enough to conclude whether or not PH$_3$ locks most of phosphorus in these solar system bodies \citep{cro2004,agu2014}.

Outside the solar system, no evidence has been found about the presence of PH$_3$ in interstellar clouds, either in the gas phase or as ice in dust grains. There is however some evidence of PH$_3$ being present in the outflows of evolved stars. A few years ago this molecule was tentatively identified in the carbon star envelope IRC~+10216. The identification was based on the $J=1-0$ rotational transition, lying at 266.9 GHz, which however appeared contaminated by a narrow line assigned to SiS in its $v=4$ vibrational state \citep{agu2008}. An emission feature attributable to the $J=1-0$ line of PH$_3$ was also observed independently by \citet{ten2008} in the spectra of IRC~+10216 and the carbon-rich envelope CRL~2688. The observation of just one line, affected by blending in the case of IRC~+10216, led to consider the detection of PH$_3$ as tentative. Here we report the detection of the $J=2-1$ rotational transition of PH$_3$ in IRC~+10216 using the HIFI instrument on board {\emph Herschel}, which definitively confirms the identification of phosphine in this source. This is the first time that PH$_3$ is unambiguously observed outside the solar system.

\section{Observations: identification of PH$_3$}

The HIFI observations of IRC~+10216 were obtained in May and from October to December 2010, in the context of two GT1 programmes dedicated to perform a line survey in all HIFI bands and to search for light hydrides at selected frequencies. Data were taken in double beam-switching mode with a spectral resolution of 1.1 MHz and a channel spacing of 0.5 MHz, and processed using the standard Herschel pipeline up to level 2, which provides fully calibrated spectra in the antenna temperature scale. The intensity scale was later on transformed to main beam brightness temperature $T_{\rm MB}$. The local oscillator was shifted in frequency to identify any emission arising from the image band. Spectra were smoothed to a spectral resolution of 1.5 MHz. For details about the data reduction we refer to \citet{cer2010}.

Phosphine was searched for in IRC~+10216 with HIFI through its $J=2-1$ and $J=3-2$ rotational transitions. This molecule is an oblate symmetric rotor and thus its rotational levels are defined by the quantum numbers $J$ and $K$, and radiative transitions are only allowed within levels of the same $K$ ladder. The $K$ ladders are grouped into two distinct forms: ortho and para, between which radiative and collisional transitions are severely forbidden. The rotational spectrum of PH$_3$ has been extensively studied in the laboratory and line frequencies are accurately known \citep{caz2006,mul2013,sou2013}. Its electric dipole moment has been measured as 0.574 D \citep{dav1971}.

\begin{figure}
\includegraphics[angle=0,width=\columnwidth]{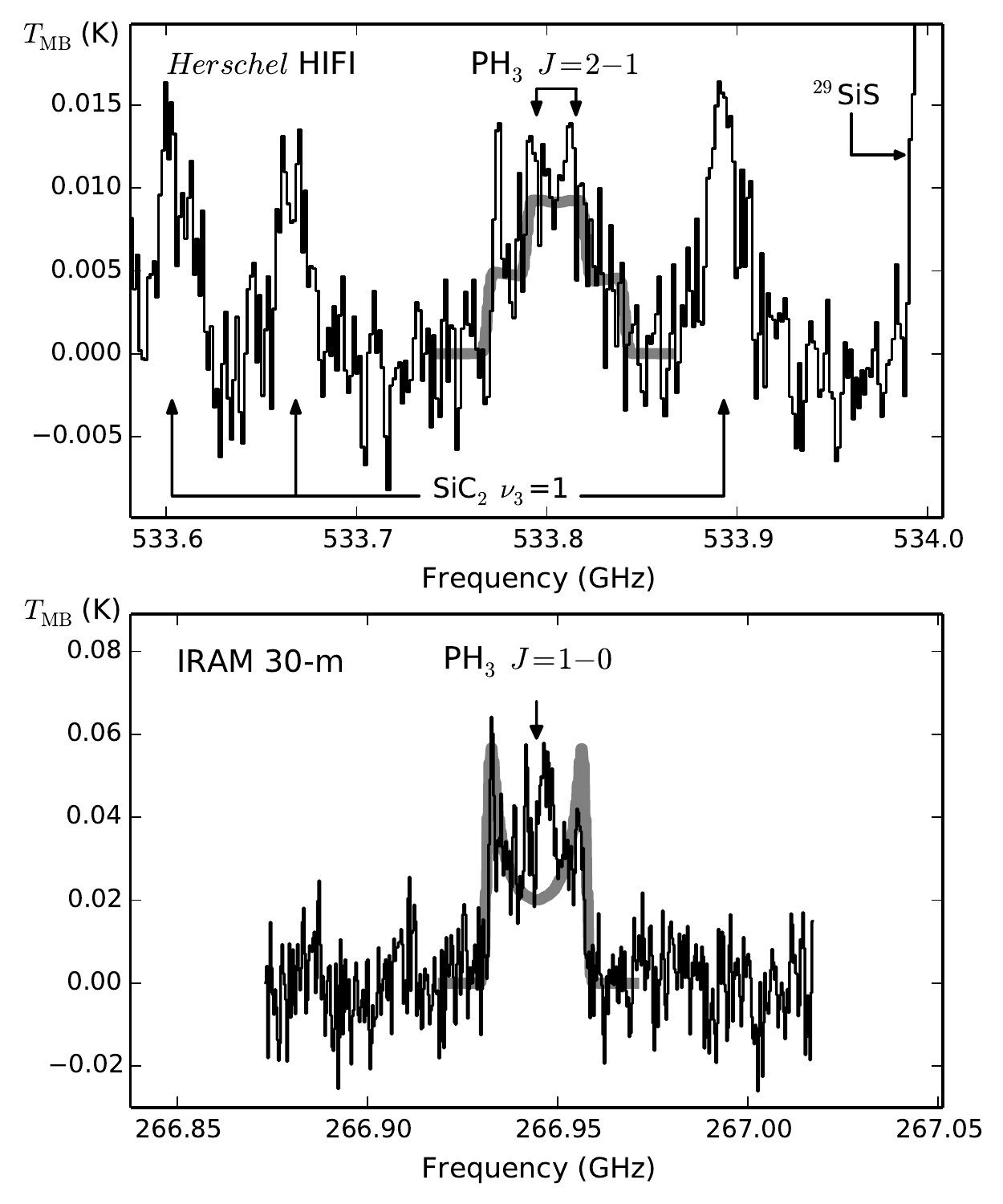}
\caption{PH$_3$ lines observed in IRC~+10216 (black histogram). The lower panel shows the $J=1-0$ transition at 266.9 GHz observed with the IRAM 30-m telescope at a spectral resolution of 0.32 MHz. An overlapping narrow line arising from SiS $v=4$ has been substracted (see \citealt{agu2008}). The upper panel shows the $J=2-1$ transition (consisting of $K=0$ and $K=1$ components) at 533.8 GHz observed with {\emph Herschel} HIFI. The spectrum, smoothed to a spectral resolution of 1.5 MHz, shows also three lines of SiC$_2$ $\nu_3=1$ and one line of $^{29}$SiS. Thick gray lines are the line profiles calculated with the LVG model (see section~\ref{sec:discussion}). The central peak observed in the $J=1-0$ line, not accounted for by the model, is probably due to a non optimal substraction of the SiS $v=4$ overlapping line.} \label{fig:lines}
\end{figure}

The $J=2-1$ rotational transition of PH$_3$ consists of two $K$ components: 2$_0-1_0$ at 533.7946 GHz and 2$_1-1_1$ at 533.8153 GHz. The HIFI spectrum of IRC~+10216 at these frequencies is shown in the upper panel of Fig.~\ref{fig:lines}. The $T_{\rm MB}$ rms noise level is 0.0024 K per 1.5 MHz channel. There is an emission feature with $T_{\rm MB}$ $\sim$0.01 K centered at 533.8 GHz which can be clearly identified with a composite of the two $K$ components of the  $J=2-1$ transition of PH$_3$. We have verified that there is no other obvious assignment for this emission feature by looking at the spectral catalog of MADEX \citep{cer2012}. Each of the two $K$ components has a linewidth consistent with an expansion velocity $v_{\rm exp}$ of 14.5 km s$^{-1}$, as most of lines arising from the outer envelope in IRC~+10216 \citep{cer2000}, including the line previously observed with the IRAM 30-m telescope and assigned to the $J=1-0$ transition of PH$_3$ (see lower panel of Fig.~\ref{fig:lines} and \citealt{agu2008}). The spectrum around 533.8 GHz also shows three narrow lines, with $v_{\rm exp}$ $\sim$8-10 km s$^{-1}$, arising from SiC$_2$ in its $\nu_3=1$ vibrational state and one from $^{29}$SiS. Lines from these species have been already observed in IRC~+10216 at millimeter wavelengths \citep{cer2000,agu2012} and are also detected across the HIFI spectral scan.

The $J=3-2$ transition of PH$_3$ lying at 800.5 GHz is not detected because the sensitivity of the HIFI spectrum at these frequencies is not good enough. The $T_{\rm MB}$ rms noise level is 20 mK per 1.5 MHz channel and this PH$_3$ line is expected to have a $T_{\rm MB}$ of some mK. Moreover, even if the sensitivity were much better, the detection of this line would be hampered by the fact that the $K=0$ and $K=1$ components, at 800.4562 GHz and 800.4871 GHz, would appear severely blended with two strong lines from SiC$_2$ (34$_{10,25}$-33$_{10,24}$ and 34$_{10,24}$-33$_{10,23}$ at 800.4842 GHz and 800.4907 GHz; \citealt{mul2012}). Intense lines from warm SiC$_2$ are numerous across the whole HIFI spectrum \citep{cer2010}. The $K=2$ component of PH$_3$, at 800.5799 GHz, does not overlap with these SiC$_2$ lines (linewidths in IRC~+10216's spectra at these frequencies are $\sim$80 MHz) and thus would be observable if the sensitivity were good enough. 

The detection of the $J=2-1$ transition of PH$_3$ with HIFI, together with the previous detection of the $J=1-0$ line \citep{agu2008,ten2008}, definitively confirms the identification of PH$_3$ in IRC~+10216. The non detection of the $J=3-2$ line in our HIFI data is consistent with the line intensities of the $J=1-0$ and $J=2-1$ lines, and thus with the identification of PH$_3$.

\section{Discussion} \label{sec:discussion}

To extract information about the abundance, spatial distribution, and excitation of PH$_3$ in the envelope of IRC~+10216 the observed lines must be confronted with the results of excitation and radiative transfer calculations. Here the calculations are based on a multi-shell LVG formalism and the model of the envelope adopted is that of \citet{agu2012}. Basically, it consists of an AGB star surrounded by a spherically expanding envelope of gas and dust. The adopted distance and mass loss rate are 130 pc and 2 $\times$ 10$^{-5}$ $M_{\odot}$ yr$^{-1}$, and dust is assumed to condense at a radius of 5 $R_*$. The stellar properties, radial profiles, and dust parameters are described in \citet{agu2012}. In the innermost layers ($<$5 $R_*$), we have adopted the downward revision of the density of particles derived by \citet{cer2013} from ALMA observations of HNC in various excited vibrational states.

Energy levels and transition frequencies of PH$_3$ in its ground vibrational state were computed from the rotational constants reported by \citet{caz2006}. Line strengths for pure rotational transitions were computed from the dipole moment measured by \citet{dav1971}. Both ortho and para forms of PH$_3$ were considered with the statistical ortho-to-para ratio of 1. Since rate coefficients for collisional excitation of PH$_3$ with H$_2$ or He are not known, we adopted those computed for NH$_3$ \citep{dan1988,mac2005}, properly corrected to the case without inversion doubling. Scaling of the rate coefficients due to the mass change from NH$_3$ to PH$_3$ is very small ($<$5 \%) and was thus not considered. No extrapolation in temperature was done, and thus the rate coefficients at 300 K, the highest temperature available in the quantum calculations of NH$_3$, were adopted at higher temperatures. The use of these rate coefficients is probably the major source of uncertainty introduced in the excitation calculations. We included rotational levels up to $J_K=7_6$ for the ortho species and $J_K=5_5$ for para PH$_3$, the highest levels involved in the quantum calculations of the collisional rate coefficients for NH$_3$. Infrared pumping to excited vibrational states was taken into account by including the first excited states of the four vibrational modes $\nu_1$, $\nu_2$, $\nu_3$, and $\nu_4$, lying at 2321, 992, 2327, and 1118 cm$^{-1}$, respectively, over the ground state. Spectroscopic constants and band intensities were taken from laboratory measurements \citep{bal1980,tar1984,fus2000,bro2002,yur2006,sou2013}. The fundamental bands of these modes lie at 4.3, 9, and 10 $\mu$m, wavelengths at which the flux in the envelope of IRC~+10216 is large \citep{cer1999}. Collisional rates for ro-vibrational transitions were assumed to be negligible compared with radiative rates.

\begin{figure}
\includegraphics[angle=0,width=\columnwidth]{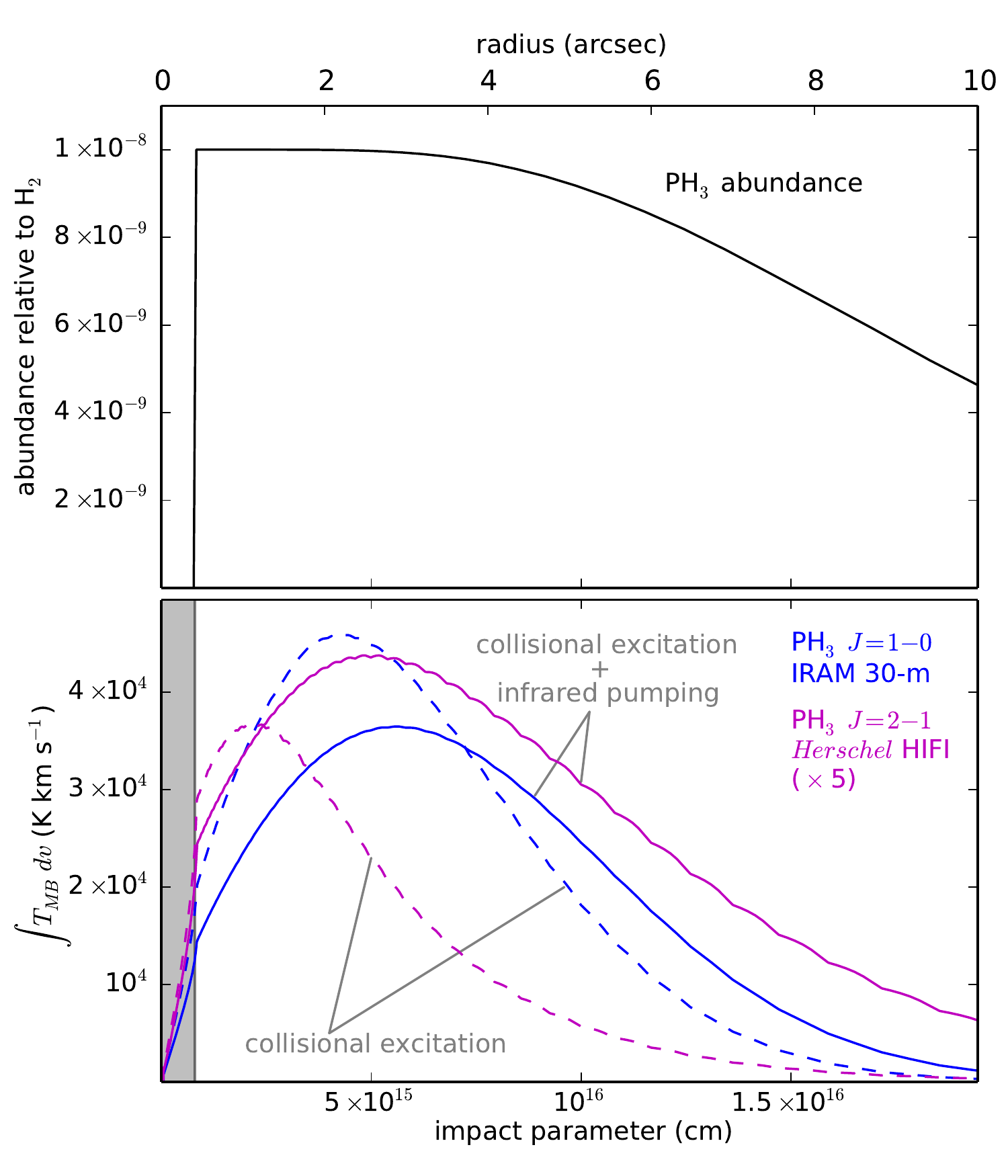}
\caption{Upper panel: fractional abundance radial profile adopted for PH$_3$. Phosphine is assumed to be present from 20 R$_*$ with an abundance of 10$^{-8}$ relative to H$_2$ and to decrease in the outer layers due to photodissociation by interstellar UV photons. Lower panel: velocity-integrated intensity computed as a function of the impact parameter for the $J=1-0$ and $J=2-1$ lines of PH$_3$. The integration over impact parameter (if expressed as angular distance in radians) of the magnitude plotted in the ordinate axis yields the calculated value of $\int$$T_{\rm MB}$$dv$. The shaded area corresponds to the region inner to 20 $R_*$. The two panels share a common abscisa axis.} \label{fig:distribution}
\end{figure}

The spatial distribution of phosphine in the envelope of IRC~+10216 is uncertain because the observed line profiles provide just limited information. In the case of the $J=1-0$ line, the substration of the blended line of SiS $v=4$ introduces an important uncertainty on the real shape of the PH$_3$ contribution. The profile of the $J=2-1$ line is complicated by the overlap of the $K=0$ and $K=1$ components and the limited signal-to-noise ratio. In any case, the PH$_3$ lines observed show a width consistent with an expansion velocity of 14.5 km s$^{-1}$, which indicates that an important part of the PH$_3$ emission comes from regions where the gas has already reached the terminal expansion velocity of the envelope. Whether phosphine is formed close to the star or farther out in the envelope is unknown, although there are reasons to suspect that it is not formed too close to the star. On the one hand the abundance of PH$_3$ predicted at thermochemical equilibrium in the hot and dense surroundings of the star is negligible, $<$10$^{-12}$ relative to H$_2$ \citep{agu2007}. On the other, infrared observations rule out the presence of an important amount of the related molecule NH$_3$ inside a radius of $\sim$20 $R_*$ \citep{kea1993,mon2000}. Following \citet{has2006} and \citet{agu2008} here we assume that PH$_3$ is formed at an inner radius of 20 $R_*$. The outer boundary is calculated with a photochemical model in which PH$_3$ is photodissociated by the external UV field using the rate guessed by \citet{mac2001}. We find that to reproduce the observed lines (see Fig.~\ref{fig:lines}) we need a PH$_3$ abundance of 10$^{-8}$ relative to H$_2$, which is just slightly higher than the value previously derived by \citet{agu2008} from the $J=1-0$ line. The adopted abundance profile is shown in the upper panel of Fig.~\ref{fig:distribution}.

Various conclusions can be drawn from a careful inspection to the excitation and radiative transfer calculations. The first is that infrared pumping to excited vibrational states plays an important role in the excitation of PH$_3$ in the envelope. It increases the excitation temperature of rotational transitions within the ground vibrational state, and therefore their line intensities, especially the $J=2-1$ and higher $J$ transitions, and extends the emitting region of the lines to larger radii. This can be appreciated in the lower panel of Fig.~\ref{fig:distribution}, where we show the contribution to the velocity-integrated line intensity of rays arising at each impact parameter. It can be seen that the bulk of the emission in the $J=1-0$ and $J=2-1$ lines comes from regions with impact parameters from the star between 1$''$ and 6$''$. That is, the emission size of the $J=1-0$ line is of the order of the main beam of the IRAM 30-m telescope (9.2$''$ at 266.9 GHz) while the $J=2-1$ line is highly diluted in the main beam of HIFI (40$''$ at 533.8 GHz). We can also appreciate that the observed lines of PH$_3$ do not trace the regions inner to 20 $R_*$. Even if we extended phosphine down to the stellar photosphere, the contribution of impact parameters shorter than 20 $R_*$ to the velocity-integrated line intensities would be less than 5 \%. In fact, our LVG calculations indicate that the observed lines are compatible with PH$_3$ being formed anywhere between the stellar surface and $\sim$100 R$_*$.

Assuming that the elemental abundance of phosphorus in IRC~+10216 is solar (P/H = 2.6 $\times$ 10$^{-7}$; \citealt{asp2009}), it turns out that HCP locks 5 \% of phosphorus \citep{agu2007,agu2012} while PH$_3$ takes 2 \%. It is likely that the rest of P is in some condensed form taking part of dust grains. In envelopes around other evolved stars gas phase molecules seem to lock a greater fraction of phosphorus than in IRC~+10216. Around half of P is in the form of HCP and PH$_3$ in the carbon-rich envelope CRL~2688 \citep{mil2008,ten2008}, while in oxygen-rich objects PO and PN contain a good fraction of the available P, around 1/4 in VY CMa \citep{ten2007,mil2008} and essentially all available phosphorus in IK Tau \citep{deb2013}.

In IRC~+10216 phosphine is formed somewhere between 1 and 100 $R_*$ from the star, although it is not clear which is the main formation mechanism. Thermochemical equilibrium calculations indicate that its abundance in the surroundings of the star is very low ($<$10$^{-12}$ relative to H$_2$) and gas phase chemical kinetics models yield no net formation in the outer layers of the envelope \citep{agu2007}. There are various plausible non-equilibrium processes at work in the inner envelope, which may be at the origin of PH$_3$ and other hydrides such as NH$_3$ and H$_2$O, for which thermochemical equilibrium predicts abundances much lower than observed. Interestingly, the NH$_3$/PH$_3$ abundance ratio in IRC~+10216 ($\sim$200; \citealt{has2006}) is similar to the solar elemental N/P abundance ratio, indicating that a similar fraction of N and P are locked in NH$_3$ and PH$_3$, respectively. A non equilibrium chemistry driven by shocks induced by the stellar pulsation has been proposed as a source of water vapour in IRC~+10216 \citep{che2011}. In this scenario, however, NH$_3$ is formed with a very low abundance and details on whether or not PH$_3$ is formed are not provided. Another mechanism proposed for the formation of hydrides such as H$_2$O and NH$_3$ in the inner layers of carbon-rich envelopes such as IRC~+10216 is photochemistry driven by the penetration of interstellar UV photons across the clumpy envelope \citep{dec2010,agu2010}. In this scenario PH$_3$ is not efficiently formed although it is likely due to the lack of relevant chemical kinetics data for phosphorus species. Another possible source of hydrides such as PH$_3$, yet to be explored, could be provided by chemical reactions taking place on dust grain surfaces. Which of these mechanims, if any, is responsible for the formation of PH$_3$ in the ejecta of evolved stars such as IRC~+10216 has yet to be investigated.

\section{Summary}

We use the HIFI instrument on board {\emph Herschel} to observe the $J=2-1$ line of PH$_3$ in IRC~+10216, which together with the previous observation of the $J=1-0$ line with the IRAM 30-m telescope, definitively confirms the first identification of PH$_3$ outside the solar system. Excitation and radiative transfer calculations indicate that the observed lines are consistent with phosphine being formed in the circumstellar envelope anywhere between 1 and 100 $R_*$ from the star with an abundance of 10$^{-8}$ relative to H$_2$. The detection of PH$_3$ challenges chemical models as no obvious formation route has for the moment been found. Prospects to put further constraints on the distribution and origin of PH$_3$ in IRC~+10216 with telescope facilities such as ALMA seem challenging. The $J=1-0$ line being blended with a SiS $v=4$ line, it makes complicated to disentangle the contribution from each line to the spatial distribution of the emission. The $J=2-1$ line at 533.8 GHz cannot be reached from ground because of the terrestrial atmospheric opacity, and the $J=3-2$ line is also severely blended with a strong line of SiC$_2$. Although PH$_3$ locks a minor fraction of phosphorus in IRC~+10216, it is together with HCP, one of the major gas phase carriers of phosphorus in the inner circumstellar layers, suggesting that it could be also an important phosphorus species in other astronomical environments. Observations of PH$_3$ in sources other than IRC~+10216, such as CRL~2688, may help to better understand its formation and the implications for the chemistry of phosphorus in space.

\acknowledgements

HIFI has been designed and built by a consortium of institutes and university departments from across Europe, Canada, and the United States (NASA) under the leadership of SRON, Netherlands Institute for Space Research, Groningen, The Netherlands, and with major contributions from Germany, France and the US. Consortium members are Canada: CSA, U. Waterloo; France: CESR, LAB, LERMA, IRAM; Germany: KOSMA, MPIfR, MPS; Ireland: NUI Maynooth; Italy: ASI, IFSI-INAF, Osservatorio Astrof\'isico di Arcetri-INAF; Netherlands: SRON, TUD; Poland: CAMK, CBK; Spain: Observatorio Astron\'omico Nacional (IGN), Centro de Astrobiolog\'ia (INTA-CSIC); Sweden: Chalmers University of Technology - MC2, RSS \& GARD, Onsala Space Observatory, Swedish National Space Board, Stockholm University - Stockholm Observatory; Switzerland: ETH Zurich, FHNW; USA: CalTech, JPL, NHSC. M.A. and J.C. thank Spanish MICINN for funding support through grant CSD2009-00038.

\end{document}